\documentclass[conference]{IEEEtran}
\IEEEoverridecommandlockouts
\usepackage{cite}
\usepackage{amsmath,amssymb,amsfonts}
\usepackage{algorithmic}
\usepackage{graphicx}
\usepackage{textcomp}
\usepackage{xcolor}
\usepackage{arydshln}

\def\BibTeX{{\rm B\kern-.05em{\sc i\kern-.025em b}\kern-.08em
    T\kern-.1667em\lower.7ex\hbox{E}\kern-.125emX}}

\usepackage[htt]{hyphenat}
\usepackage{multirow}

\begin{document}

\title{NorBERT: NetwOrk Representations through BERT\\ for Network Analysis \& Management}

\author{\IEEEauthorblockN{Franck Le, Davis Wertheimer, Seraphin Calo, Erich Nahum}
fle@us.ibm.com,  davis.wertheimer@ibm.com, scalo@us.ibm.com, nahum@us.ibm.com \\
IBM Research US 
 
}


\maketitle

\begin{abstract}
Deep neural network models have been very successfully applied to  Natural Language Processing (NLP) and Image based tasks. Their application to network analysis and management tasks is just recently being pursued. Our interest is in producing deep models that can be effectively generalized to perform well on multiple network tasks in different environments. A major challenge is that traditional deep models often rely on categorical features, but cannot handle unseen categorical values. One method for dealing with such problems is to learn contextual embeddings for categorical variables used by deep networks to improve their performance. In this paper, we adapt the NLP pre-training technique and associated deep model BERT to learn semantically meaningful numerical representations (embeddings) for Fully Qualified Domain Names (FQDNs) used in communication networks. We show through a series of experiments that such an approach can be used to generate models that maintain their effectiveness when applied to environments other than the one in which they were trained.
\end{abstract}

\section{Introduction}
\label{sec:intro}

\textit{How to improve the generalization performance of deep models for network analysis and management tasks?} As deep models are gradually being adopted for a wide range of network operations (e.g., packet classification~\cite{8713803, lotfollahi2020deep, 10.1145/3341302.3342221}, device classification~\cite{unsw, 9151339, 8440758}, resource management~\cite{10.1145/3005745.3005750}, congestion control~\cite{10.1145/2486001.2486020, 8790762, 8668690, 10.1145/3512798.3512815}, routing~\cite{NIPS1993_4ea06fbc, 8851733, 8701570}, and scheduling policies identification~\cite{9614298}), this question becomes ever more salient. Although the generalization performance of deep models is a general problem, computer and telecommunication systems present unique challenges that exacerbate the issue. For example, local network configurations, geographic load balancing, or software updates can all cause deep models to overfit the training data, and perform poorly in the relevant outside environments. 

\textbf{Motivation:} To illustrate these problems, we consider the task of detecting and classifying devices (e.g., manufacturer, operating system) present in a network from passive network monitoring. Because Internet of Things (IoT) devices are particularly vulnerable to attack, due to their poor security~\cite{10.1145/2834050.2834095}, network administrators may want to identify devices from specific manufacturers or types (e.g., cameras) to apply different policies (e.g., quarantine, network isolation)~\cite{unsw}. However, while connections to a server ``\texttt{server-54.us-east-1.update.xyz.com}'' may reveal the presence of an IoT device from the manufacturer \texttt{xyz}~\cite{9151339}, that same device may instead connect to a different server ``\texttt{server-27.us-west-2.update.xyz.com}'' when deployed in a different enterprise network because of geographic load balancing.
Network administrators would immediately recognize that the device is accessing the same service, but through a server
in a different data center and location for better performance. However, to a deep neural network, the fully qualified domain name (FQDN) ``\texttt{server-27.us-west-2.update.xyz.com}'' is instead a new unseen value. 
Our deep device classifier may therefore perform poorly on the new enterprise network. 

Similarly, devices running distinct software versions in different enterprise networks may cause deep models to perform well in the training environment, but poorly in new settings. As an example, devices of interest may connect to the server ``\texttt{fw-update.xyz.com}'' in the training setting, but devices from the same manufacturer in other enterprise networks may be running newer software versions, and instead connect to ``\texttt{fw-update2.xyz.com}''. As in the prior example, the FQDN ``\texttt{fw-update2.xyz.com}'' is an entirely unseen value to our deep model, leading to poor performance in new settings.

To quantify the performance degradation, we train several Gated Recurrent Unit (GRU) models to predict the type (e.g., tv, refrigerator, hub, camera, doorbell) and manufacturer (e.g., Amazon, Apple, Belkin, Canary, D-link) of IoT devices, by analyzing their DNS traffic\footnote{Details of the model and training dataset are provided in later sections.}. All the models exhibit poor generalization performance. For example, the device type classifier achieves a weighted average F1-score of 0.997 across all classes on the validation dataset, defined as 25\% of the entire training set (traditional train-test split validation approach), but its performance drops to 0.592 when applied to a different publicly available network trace~\cite{unsw}.

\textbf{Research problem:} Although the above examples demonstrate the problem in the context of device classification, the root cause is more general, and affects many other network analysis and management tasks. More specifically, \textit{traditional deep models' approaches often rely on categorical features, but cannot handle unseen categorical values}: performance degradation stems from an inability to learn the semantic meanings of new categorical values. Instead, if deep models were able to learn that the newly observed FQDN ``\texttt{server-27.us-west-2.update.xyz.com}'' is actually semantically similar to ``\texttt{server-54.us-east-1.update.xyz.com}'', they may generalize better. Even in other tasks not relying on FQDNs, relevant features may still be categorical in general, and so the capability to learn semantic meaning of newly encountered values is essential to generalization performance. For example, for security vulnerability analysis, resource management (e.g., scaling of Virtualized Network Functions), or congestion control/routing, respectively, deep models may take into account the SSL cipher suite~\cite{wagner1996analysis} of connections, the cloud instance types/flavors~\cite{ou2012exploiting}, and the QoS or Explicit Congestion Notification (ECN) codes~\cite{10.1145/78952.78955} in IP packets, respectively, to determine the appropriate actions. But how can these models learn the semantic meanings of newly encountered values, not observed during training?

\textbf{Contributions:} In this paper, we first adapt a Natural Language Processing (NLP) pre-training technique and associated deep model called Bidirectional Encoder Representations from Transformers (BERT)~\cite{devlin2018bert}, to learn semantically meaningful numerical representations, also called embeddings, for FQDNs. More specifically, we modify the tokenization step: BERT relies on WordPiece tokenization to identify basic word components -- e.g., splitting ``cars'' into ``car'' and ``s'' -- and determine the tokens for which to learn embeddings. 
Instead, we exploit the hierarchy of domains, which was designed to manage and resolve domains, to define the individual tokens.

Second, we demonstrate that applying the above algorithm allows embeddings to capture the semantic information behind FQDNs, and significantly improves deep models' generalization performance. Continuing the previous example, the performance of the GRU-based IoT device type predictor improves from 0.592 to 0.965 when applied to the same independent dataset, but with embeddings learned through the proposed algorithm. 
We also compare performance when using context independent embeddings (e.g., GloVe~\cite{pennington-etal-2014-glove}): with context independent embeddings, the performance of the downstream model on the independent dataset remains poor at 0.585. These experiments demonstrate the benefits of using BERT and deriving contextual embeddings for FQDNs. 

Third, we further analyze the learned embeddings, and present (unsupervised) tasks network administrators could perform thanks to these novel embeddings (e.g., analysis of user behaviors). Finally, we discuss approaches for learning embeddings that capture semantic relationships for network categorical values, other than FQDNs; and present preliminary results. In particular, we found that the proposed approach can find meaningful relationships: For example, HTTP and HTTPS services are discovered to be semantically very similar; and strong TLS ciphersuites that rely on the same set of algorithms, and differ only in the keys' lengths are also discovered to be close. We conclude the paper with a discussion of future research directions.
\section{Background: Word Embeddings in NLP}
\label{sec:background}

This section provides a brief overview on the motivation, challenges, and techniques for creating word embeddings in NLP. 
Word embeddings are needed because deep neural networks cannot handle strings, only numerical vectors. As such, converting text to vectorial representations in high-dimensional latent spaces has been an active research area. 

Although one hot encoding provides a simple numerical representation for categorical values, they do not capture any semantic information between words. Instead, every pair of words is equidistant in the latent space.

\textbf{Word2Vec}: In 2013, researchers developed Word2Vec~\cite{mikolov2013efficient}, a popular technique to learn numerical representations for words that captures their semantic meaning via proximity. Two words that are close in meaning (e.g., ``king'', and ``queen'') would have word embeddings (numerical vectors) that are close in distance (e.g., cosine similarity). To achieve this, Word2Vec relies on two neural network variants that calculate word embeddings based on the words’ context: Continuous Bag-of-Words (CBOW) predicts the current word based on the context; and Skip-gram instead predicts the closely related context words to an input word.

\textbf{GloVe:} In 2014, researchers presented Global Vectors for Word Representation (GloVe)~\cite{pennington-etal-2014-glove}, an extension of Word2Vec that capture global contextual information in a text corpus. More specifically, while Word2Vec ignores the frequency of co-occurrence of words (i.e., the number of times
word $j$ occurs in the context of word $i$), GloVe computes word embeddings taking co-occurrence into account. For this reason, GloVe embeddings are often considered a representation of the training corpus in lower dimensionality, and which reflects the word-word contextual relationships.

\textbf{BERT:} More recently, in 2018, researchers introduced Bidirectional Encoder Representations from Transformers (BERT)~\cite{devlin2018bert} to compute contextual embeddings. 
Word2Vec and GloVe compute context independent embeddings: given a word, its numerical representation is the same regardless of its position in a sentence, or the different meanings it may have. For example, in the sentence ``\textit{Alice had a picnic by the river bank, and then went to the bank to open an account}'', the two occurrences of the word ``\textit{bank}'' have the same embedding. In contrast, with BERT, the two occurrences of the word ``\textit{bank}'' have different embeddings.
To achieve this, a tokenizer~\cite{6289079, sennrich-etal-2016-neural} splits the text into words or subwords (e.g., ``\textit{banks}'' would be decomposed into ``\textit{bank}'' and ``\textit{s}''). Then, BERT is trained on two unsupervised tasks:

\noindent $\bullet$ Masked LM: A fraction of the words in the text are replaced with a [MASK] token, and the model's goal is to predict the original value of the masked words. However, compared to previous solutions which read the text input sequentially (e.g., left-to-right, or right-to-left), BERT applies the bidirectional training of a Transformer~\cite{NIPS2017_3f5ee243}, an attention model~\cite{bahdanau2014neural}, to language modelling. 
This allows the model to learn the context of a word based on all of its surroundings.

\noindent $\bullet$ Next Sentence Prediction: Given two sentences, the model is to predict whether the second sentence is the next sentence to the first one in the initial text, or if the second sentence has been randomly sampled.

BERT set new state-of-the-art results for a wide range of natural language processing tasks (e.g., text classification, question answering), and a large number of variants~\cite{lan2019albert, liu2019roberta} have since been proposed. 

\section{Contextual Embeddings for Data Networks}

Our goal consists of learning contextual embeddings (i.e., numerical representations that capture semantic relationships within a sequence) for categorical variables used by deep networks trained for network analysis and management tasks, to improve their performance. 

\subsection{What Categorical Variable?}

In this paper, we focus on fully qualified domain names (FQDN), the complete, human-readable domain name (e.g., \texttt{mascots.iitis.pl}) for servers on the Internet. This is because FQDNs are semantically rich, and either explicitly, or implicitly, carry a lot of information. For example, in the FQDN ``\texttt{x.y.z}'', the Top Level Domain (TLD) ``\texttt{z}'' distinguishes commercial enterprises (``\texttt{.com}''), government entities (``\texttt{.gov}''), educational institutions (``\texttt{.edu}''), and nonprofit organizations (``\texttt{.org}''); or can also indicate the country where the domain is hosted: ``\texttt{.ca}'' for Canada, ``\texttt{.uk}'' the United Kingdom, etc. The Second Level Domain (SLD) ``\texttt{y}'' commonly indicates the organization that registered the domain name with a registrar. Subsequent domains may indicate services, department, or other structures in that organization.

Given their semantic richness, a large number of  solutions rely on the analysis of FQDNs to detect botnets, phishing, spam, abnormal activities, and for classification~\cite{10.5555/2028067.2028094, 10.1145/2584679, 9151339, 10.5555/2460416.2460425, Holz:2008, Perdisci:2009, 10.1145/1879141.1879148}. As such, we set out to investigate whether recent techniques developed in NLP can capture this information, and derive semantically meaningful embeddings for FQDNs.

\subsection{How To Tokenize FQDNs?}
\label{sec:tokenize}

Studies have revealed that tokenizers play a critical role in the semantic relations that NLP models learn from a corpus~\cite{singh-etal-2019-bert}. Because FQDNs have their own syntax, they demand a custom tokenization approach. 

We propose to exploit the hierarchical structure of FQDNs.
While FQDNs may simply appear as strings of numbers, letters and hyphens, with a maximal length of 255 characters, they actually embed a hierarchical structure, where the hierarchy levels of an FQDN are read from right to left, with full stops acting as delimiters between the different levels. As such, we truncate each FQDN to the $k$-th level, where $k$ is a hyper-parameter that allows one to control the vocabulary size. More precisely, we replace every FQDN, \texttt{FQDN},  with the corresponding token:  
\texttt{token = ".".join(FQDN.split(".")[:min(k, len(FQDN.split(".")))])}

We leave the definition, and comparison, of more advanced tokenizers for FQDN (e.g., methods that may give larger weights to higher hierarchical levels) for future work.

\subsection{How To Contextualize the Embeddings?}
\label{sec:norbert}

Neural language models have made great strides in language understanding through their ability to contextualize the tokens from a given input sequence. For example, in the sentence, ``I visited the river bank, then went to the bank to make a withdrawal,'' the token ``bank'' takes on two entirely separate meanings based on context: the first refers to a geographical feature, while the second to a financial institution. A straightforward mapping of single tokens to single embeddings will struggle to reflect this semantic dual nature. An encoder-based neural language model restores this ability, however, by remapping input embeddings to better reflect the context of the input sequence - e.g., the embedding for the first ``bank'' may get shifted closer to that for ``shore'', while the second may move closer to ``ATM''. 
By learning and remapping the input embeddings, the language model gains the ability to extract high-level semantic information from a given input sequence, which has proven useful for many downstream tasks. 

In NorBERT, we propose a similar contextualization/remapping operation to capture the high-level semantics of network traffic sequences, using a small BERT-based model. Our training task is the standard Masked Language Modeling task, implemented as in Roberta~\cite{liu2019roberta}: given an input sequence, $15\%$ of tokens are randomly chosen as training anchors. $80\%$ of those anchors are replaced with a standard masking token, and the model must fill in the blanks from its learned vocabulary. A separate $10\%$ of anchors are left unchanged, to encourage the model to preserve the meanings of unaltered non-anchor tokens, and the remaining $10\%$ are replaced with a random token from the vocabulary, to encourage the model to contextualize all tokens in the sequence, rather than only those that are masked off. We use the Adam optimizer~\cite{kingma2014adam} and a linear annealing schedule for learning rate. 
Training sequence length ranges from 8 to 64, resampled every batch. Further hyperparameters and architecture details are provided in Table~\ref{table:bertvars}. 

\begin{table}[t]
\centering
\begin{tabular}{|c|c|c|}
\hline
Variable & Description & Value \\
\hline
$D$ & Dimension of embeddings & 128\\
$N$ & Number of successive FQDNs in a sequence & [8,16,32,64] \\
$k$ & Truncation level in FQDN hierarchy (Sec.~\ref{sec:tokenize})& 3 \\
$h$ & Number of BERT self-attention heads & 8\\
$b$ & BERT batch size & 32\\
$l$ & Number of attention layers & 4\\
$l_r$ & Learning rate & 0.001\\
$w$ & Weight decay & 0.01\\
\hline
\end{tabular}
\smallskip
\caption{Summary of BERT hyper-parameters and variables}
\label{table:bertvars}
\end{table}
\section{Experiments}
We conduct a number of experiments to evaluate the benefits of contextual embeddings for network analysis. We first present the datasets, then describe the network tasks, and finally discuss the results.

\subsection{Datasets}
\label{sec:datasets}

The datasets consist of packet capture (pcap) from three independent environments:

\noindent $\bullet$ Smart Lab - UNSW: A lab with 21 IoT devices, including cameras, switches and triggers, hubs, air quality sensors, electronics, healthcare devices, and light bulbs was setup~\cite{unsw}, and all traffic was collected, for two weeks in 2016, using \texttt{tcpdump}. The devices are labeled with the manufacturers and device types (e.g., Amazon Echo, Netatmo Welcome, etc.)

\noindent $\bullet$ Smart Lab - Private: Instrumented very similarly to the previous one, a private lab comprises 65 IoT devices from 31 vendors (e.g., Amazon, Apple), and 24 different types (e.g., TV, bulbs, camera, doorbell, hub, refrigerator, thermostat, etc.) The devices are also labeled with their manufacturer, and type. We focus on their traffic captured over 31 days in 2017.
    
\noindent $\bullet$ Large enterprise network: Traffic from an enterprise network located in North America, and which comprises slightly more than 3,000 hosts including servers, laptops, and phones, are captured at the border routers with the Internet for several 24 hour periods. The devices are not labeled.

We preprocess each dataset by applying Zeek~\cite{Bro}, a network analysis framework that parses packet captures and extracts fields of interest. We focus on the DNS traffic, and define an instance as a sequence of $N$ successive FQDNs queried by a same device. We label the sequences with the type and manufacturer of the IoT devices that originated them.

\subsection{Methodology}
\label{sec:methodology}

\textbf{Tasks:} We consider the tasks of predicting the (1) IoT device type (e.g., camera, doorbell, hub), and (2) their manufacturer (e.g., Amazon, Apple, Belkin) from passive network monitoring. More specifically, after observing $N$ (e.g., 32)  successive DNS queries that a device queried, can a model correctly predict its device type and manufacturer?
We trained gated recurrent units (GRU) models~\cite{cho-etal-2014-learning}, which are a popular variant of Recurrent Neural Networks architectures, and often considered an improvement to Long Short-Term Memory (LSTM) networks. Details of the model architecture,  hyper-parameters, and other variables are provided in Table~\ref{table:variables}.

\begin{table}[t]
\centering
\begin{tabular}{|c|c|c|}
\hline
Variable & Description & Value \\
\hline
\hline
$D$ & Dimension of embeddings & 128\\
$N$ & Number of successive FQDNs in a sequence & 32 \\
$k$ & Truncation level in FQDN hierarchy (Sec.~\ref{sec:tokenize})& 3 \\
$h$ & GRU Number of features in the hidden state & 64\\
$b$ & GRU Batch size & 1024\\
$l$ & GRU Number of recurrent layers & 2\\
$l_r$ & Learning rate & 0.001\\
\hline
\end{tabular}
\smallskip
\caption{Summary of classifier hyper-parameters and variables}
\label{table:variables}
\end{table}
\begin{table*}[t]
\small
\centering
\begin{tabular}{|c|c|c|c|}
\hline
\multirow{2}{*}{Task} & \multirow{2}{*}{Embedding} &  Training Dataset & Independent\\

 &  &   Testing Subset (Training:Testing split) &  Validation Dataset\\ 
\hline
 & Random & 0.997 & 0.592 \\
Device Type  & GloVe & 0.994 &  0.585 \\
 & NorBERT & 0.998 & \textbf{0.965} \\
\hline
 & Random & 0.996 & 0.588 \\
Manufacturer  & GloVe & 0.998 & 0.726 \\
 & NorBERT & 0.981 & \textbf{0.906} \\
\hline
\end{tabular}
\smallskip
\caption{Average Weighted F-1 Scores}
\label{table:results}
\end{table*}
\textbf{Performance metrics:} To evaluate the models' generalization performance, we first train the models on the larger labeled dataset, i.e., Smart Lab - Private. We focus on the classes that have a large enough support (i.e., number of instances is larger than threshold). We split the entire dataset into training and testing subsets along a 75:25 ratio, and analyze the weighted average F-1 score on the test subset to avoid overfitting. Then, we apply the trained models on the other independent labeled dataset, i.e., Smart Lab - UNSW. Because the two datasets have a number of common, but also different classes, we only consider the instances from Smart Lab - UNSW that belong to classes seen during the training phase\footnote{Methods have recently been developed for this problem (e.g.,~\cite{10.5555/3045390.3045502, 10.5555/3295222.3295387, liang2017enhancing, 10.5555/3327757.3327819}),
but we leave the identification of out of distribution instances for future work.}. We finally compute, and compare, the average weighted F-1 score on the independent dataset.

\textbf{Embeddings:} To convert FQDNs, into numerical vector representations to be processed by neural networks, we compute and compare embeddings using three approaches. However, before describing each of them in detail, we present the training corpus, and the common preprocessing steps.


We first create a training corpus from the sequences of successive FQDNs queried by the same device, from all three datasets (Section~\ref{sec:datasets}). Learning embeddings does not require labels, and as such, the trace from the large enterprise network can also be exploited. 
Then, we preprocess the training corpus by filtering out domains that are of little value. For example, we exclude FQDNs that end with ``\texttt{in-addr.arpa}'', which are mainly used for reverse DNS. This is similar to removing stop words in natural language processing. 
In addition, to reduce the vocabulary size, and allow a fair comparison across the methods, we truncate the FQDNs to the same level (Section~\ref{sec:tokenize}) with $k$ set to 3. 
%
Finally, we implement and compare the following approaches for computing embeddings:

    \noindent $\bullet$ \textit{Random}: A simple and common approach consists in initializing the embedding to random values, which are then adjusted 
    through backpropagation.
    
    \noindent $\bullet$ \textit{GloVe}: We learn context independent embeddings for the FQDNs through GloVe (Section~\ref{sec:background}). 
    
    \noindent $\bullet$ \textit{NorBERT}: We train an FQDN embedding contextualization model using the procedure outlined in Section~\ref{sec:norbert}. 

\subsection{Results}
\label{sec:results}

The results are summarized in Table~\ref{table:results}. First, we observe that with \textit{randomly initialized} embeddings, the performance of the models -- both for predicting the device type, and the manufacturer -- suffer significant drops from F-1 scores larger than 0.996 on the training dataset, to scores lower than 0.588 on the independent validation dataset. Those numbers confirm the model generalization problem for network analysis tasks (Section~\ref{sec:intro}). 
Second, we note that \textit{context independent} embedding learnt through GloVe may improve the performance. For example, the model to predict the device's manufacturer increased its F-1 score on the independent validation dataset from 0.588 to 0.726 thanks to context independent embeddings.
Most importantly, \textit{contextual} embeddings computed through the proposed NorBERT approach allow models to outperform other approaches. The average weighted F-1 scores of the models to predict device type and manufacturer reach 0.965 and 0.906, respectively. These large improvements demonstrate the benefits of \textit{contextual} embeddings, and their ability to increase the models' generalization performance.
\section{Further analysis of embeddings}

In this Section, we further analyze the learned embeddings. For example, for a given FQDN, we can retrieve the nearest neighbor. In NLP, nearest neighbor search has been an important building block for real world applications such as text search and recommender systems. We show how such analysis in the context of FQDNs can reveal new insights, and enable new unsupervised  analysis tasks.

\subsection{Nearest Neighbor}
\label{sec:nn}

To verify the quality of word embeddings, NLP studies often look at words and their most similar words. For example, the word ``France'' should intuitively be similar to ``Spain'', or some other countries. We first conduct a similar analysis, using the learned embeddings from our NorBERT model. We focus on the packet capture from the large enterprise network. Given a list of FQDNs, we retrieve their nearest neighbor using the cosine similarity to compute the distance between any pair of embeddings. 

\begin{table}[h]
\small
\centering
\begin{tabular}{|c|c|}
\hline
FQDN & Nearest Neighbor \\
\hline
 www.theatlantic.com & www.wired.com \\
 www.theguardian.com & www.wired.com \\
 getpocket.com & www.newyorker.com \\
 www.gq.com & www.buzzfeednews.com \\
 www.fastcompany.com & www.wired.com \\
 www.harpersbazaar.com & www.wired.com \\
 www.theringer.com & lifehacker.com \\
 www.wired.com & www.nytimes.com \\
 fivethirtyeight.com & www.vox.com \\
 www.politico.com & www.wired.com \\
\hline
 ctldl.windowsupdate.com & www.bing.com \\
 activity.windows.com & edge.microsoft.com \\
 content.office.net & time.windows.com \\
\hline
 dscg.akamaiedge.net & dscj.akamaiedge.net \\
\hline
 nvidia.github.io & download.nvidia.com \\
 \hline
 desktop-\#\#\#\#\#.\#\#\#\#\#.com & win-\#\#\#\#\#.\#\#\#\#\#.com \\
 desktop-\#\#\#\#\#.\#\#\#\#\#.com & laptop-\#\#\#\#\#.\#\#\#\#\#.com \\
 laptop-\#\#\#\#\#.\#\#\#\#\#.com & laptop-\#\#\#\#\#.\#\#\#\#\#.com \\
\hline

\end{tabular}
\smallskip
\caption{Excerpt of Nearest Neighbors}
\label{table:nn}
\end{table}

Table~\ref{table:nn} presents an excerpt of the results. We observe
that the first set of FQDNs and their nearest neighbors do appear similar: They are American or English websites with articles in politics, foreign affairs, business, culture, or technology. Similarly, the next sets of FQDNs may have different second level domains (SLD) (e.g., windows, bing, windowsupdate), or third level domain, but they all appear to be owned by the same organization. For the last set of FQDNs, they all point to computers in the enterprise network. The second level domain has been anonymized with a fixed number of characters ``\#''. Network administrators in that enterprise assign a default hostname to computers starting with either ``desktop-'', ``laptop-'', or ``win-'' and followed by a unique identifier, also anonymized. However, for many such computers, their nearest neighbor seems to be another computer from the same domain. These results confirm the embeddings capture semantic relationships between FQDNs, and the latest observation suggests that computers in that enterprise may form clusters, and could point to outliers. 

\subsection{Unsupervised Learning}

\begin{figure}[t]
\centering
\includegraphics[width=1.0\columnwidth]{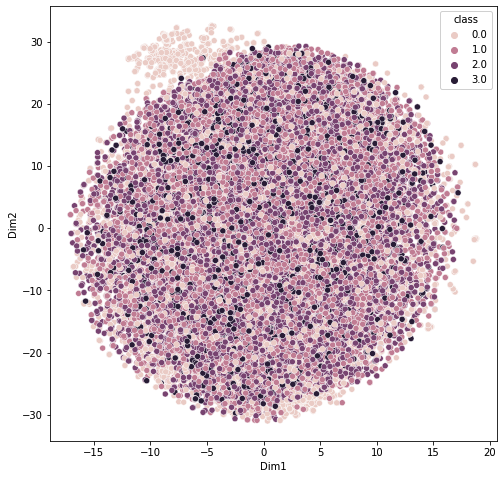}
\caption{Projection of the FQDN embeddings from the large enterprise network into a two-dimensional plane using t-SNE}
\label{fig:tsne}
\end{figure}  

Insights from the previous analysis (Section~\ref{sec:nn}) suggest that FQDNs with similar semantics may cluster together, and can help unearth patterns.
As such, we explore and  visualize the data through projection: we project the high dimensionality embeddings (128 dimensions) into a two-dimensional plane using t-distributed Stochastic Neighbor Embedding (t-SNE)~\cite{NIPS2002_6150ccc6}, a nonlinear dimensionality reduction algorithm. Figure~\ref{fig:tsne} shows the result. A point in the graph represents a FQDN. We expect users' computers to have similar embeddings. Any user's computer that differs significantly may indicate an unexpected behavior (e.g., misconfiguration). To detect any such endpoint, we distinguish FQDNs starting with the prefix ``laptop-'', ``desktop-'', and ``win-'' respectively, and ending with the second level domain of that enterprise with a different color, and the label 1, 2, and 3, respectively.

\begin{table}[t]
\small
\centering
\begin{tabular}{|c|c|c|}
\hline
FQDN & Dim1 & Dim2 \\
\hline
mirrors.wcupa.edu & -8.886705 & 31.27709 \\
centos-distro.1gservers.com & -7.9791565 & 31.273209 \\
mirror.metrocast.net & -7.9935045 & 31.254797 \\
mirrors.clouvider.net & -8.053765 & 31.230501 \\
centos.vwtonline.net & -8.039425 & 31.224531 \\
ftpmirror.your.org & -8.013818 & 31.222664 \\
us.oneandone.net & -8.051469 & 31.2198 \\
mirror.jaleco.com & -8.023687 & 31.219189 \\
repos.eggycrew.com & -8.040249 & 31.210453 \\
mirrors.usinternet.com & -7.978715 & 31.20442 \\
registry-1.docker.io & -7.223346 & 31.19538 \\
auth.docker.io & -7.210015 & 31.194843 \\
mirrors.hoobly.com & -8.009627 & 31.185928 \\
\hline

\end{tabular}
\smallskip
\caption{Excerpt of FQDNs, and their coordinates in the two-dimensional projection}
\label{table:mirrors}
\end{table}

We observe that the FQDNs starting with the prefix ``laptop-'', ``desktop-'', and ``win-'' , and ending with the second level domain of that enterprise, actually form a compact cluster with no outstanding outlier -- a desirable property for network administrators. In contrast, some embeddings, in the upper left corner of the figure appear to protrude from the main cluster. Table~\ref{table:mirrors} presents an excerpt of those FQDNs. Their names indicate a majority of them correspond to mirror servers, and repositories. Intuitively, those servers behave differently from users' computers.
These results confirm that exploration of FQDN embeddings can reveal new patterns and insights.

In addition to visual inspection, a more systematic approach to explore the data and discover patterns could consist in applying hierarchical clustering. However, due to space limitations, we leave this for future work. 

\subsection{From Token to Sequence}

A major characteristic of contextual embeddings lies in their ability to derive numerical representations whose values can change based on the context. As such, analyzing sequences of FQDNs, rather than standalone FQDNs, would further exploit their strength. For example, one can passively monitor the DNS queries submitted by each host, create sequences of $N$ FQDNs, and compute their embeddings. Then, analysis of the sequence embeddings, either through projection and visual inspection, or through hierarchical clustering, could reveal abnormal user behaviors.

\begin{figure}[t]
\centering
\includegraphics[width=1.0\columnwidth]{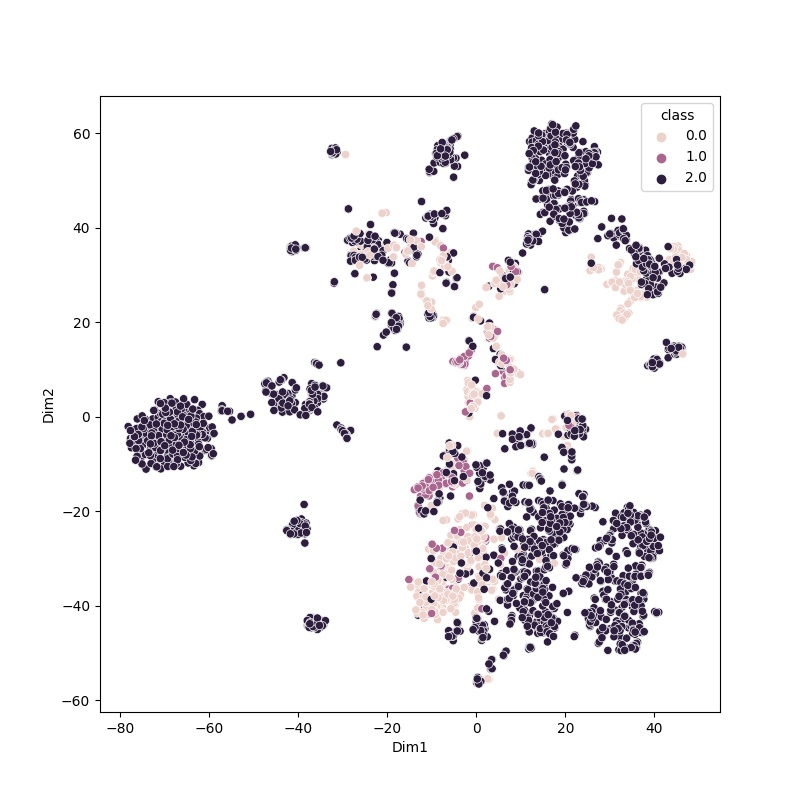}
\caption{Projection of the sequence embeddings from the large enterprise network into a two-dimensional plane using t-SNE}
\label{fig:tsne-sequence}
\end{figure}

To illustrate this, we consider a different 24 hour trace from the large enterprise network than the one that was used in Section~\ref{sec:methodology}. We define a sequence as the list of $N$ successive FQDNs queried by a single host. This sequence is fed into NorBERT to compute the contextualized embedding for each FQDN. We then derive the sequence embedding as the mean of the $N$ contextual embeddings. Finally, we again project the sequence embeddings into a two-dimensional plane using t-distributed Stochastic Neighbor Embedding (t-SNE). Figure~\ref{fig:tsne-sequence} shows the result. Each dot represents the embedding of a sequence of $N$ FQDNs queried by a single host. If the hostname of that endpoint includes specific sub-strings indicating the host is a desktop or a laptop, the dot is marked with a value of 0 or 1, respectively. All other hosts are marked with a value of 2. Network administrators can then look more closely at the outliers. An outlier indicates a sequence of FQDN that is considerably different from those typically observed by other members of a class (e.g., laptop, desktop). 
Note that compared to the uncontextualized FQDN embeddings in Fig.~\ref{fig:tsne}, the contextualized host sequence embeddings are far more distinctively clustered and semantically rich, indicating that NorBERT truly is contextualizing meaningfully, and performing non-trivial network traffic analysis, despite the lack of any human-annotated labels or guidance during the training process. 

\section{Going Beyond FQDN}

This Section discusses approaches to, and presents preliminary results on, learning contextual embeddings for categorical variables other than FQDN. While previous sections have shown how embeddings for FQDN can be computed, and then used for different network analysis tasks, one may question: Can we compute embeddings for other network categorical variables? For example, as the destination port represents the server application, destination port 80 (HTTP~\cite{rfc2616}) may be considered semantically closer to port 5684 (CoAP~\cite{rfc7252}) than port 22 (SSH~\cite{rfc4254}). Similarly, as the ciphersuite field in the TLS handshake~\cite{rfc5246} represents the set of algorithms used to secure connections, some ciphersuites may be considered semantically closer (e.g., strong versus weak). Can we learn embeddings for network categorical variables that capture such information?

One major challenge to learning embeddings consists in defining the context, and the tokens. For example, network traffic is often viewed as a sequence of bytes. In such a case, what would the delimiters and tokens be? In natural language processing, white space and punctuation are commonly used as  delimiters to extract tokens. What are the equivalent of white space and punctuation for network data?

\begin{table}[t]
\small
\centering
\begin{tabular}{|c|c|}
\multicolumn{2}{c}{Packet byte view}\\
\hline
\multicolumn{2}{|c|}{14 cc 20 51 33 ea 30 8c fb 2f ea b2 08 00 45 00}\\
\multicolumn{2}{|c|}{00 83 d2 9e 40 00 40 06 7f c2 c0 a8 01 6a 34 57}\\
\multicolumn{2}{|c|}{f1 9f 9f 3f 01 bb 05 70 2c 25 28 86 ab 7d 80 18}\\
\hline
\multicolumn{2}{c}{Protocol level view}\\
\hline
Protocol field & Protocol value\\
\hline
Ethernet II Destination & 14 cc 20 51 33 ea\\
Ethernet II Source & 30 8c fb 2f ea b2\\
Type & IPv4 (08 00)\\
\hdashline
IP Verion & 4 (4) \\
IP Header Length & 20 (5) \\
DiffServ & 00 (00)\\
IP Total Length & 142 (00 8e) \\
IP Identification & d2 9e\\
Flags & 02 (Don't fragment) (40)\\
... & ...\\
\hline
\end{tabular}
\smallskip
\caption{Different views of network data}
\label{table:wireshark}
\end{table}

\begin{table}[t]
\small
\centering
\begin{tabular}{|c|}
\hline
(Ethernet II Destination) (14 cc 20 51 33 ea)\\
(Ethernet II Source) (30 8c fb 2f ea b2) (Type) (IPv4) \\
(IP Verion) (4) ... \\
\hline
\end{tabular}
\smallskip
\caption{Representation of network data}
\label{table:representation}
\end{table}

To address those challenges, we propose to consider network data at a higher level than sequences of bytes. We argue that bytes are in reality mainly the implementation of protocols, and the exchange of information, between clients and servers. As such, network data can be treated as sequences of (protocol field, protocol value) pairs. As illustrated in Table~\ref{table:wireshark}, network traffic analyzers (e.g., \cite{10.5555/1202316}) rely on protocol parsers to extract protocol fields and values, and present to users who can then analyze the data more easily. A token can then be either a protocol field, or protocol value; and the equivalent of a sentence from natural language processing can be a sequence of tokens as depicted in Table~\ref{table:representation}.

\textbf{Experiments:} To evaluate what information network embeddings learned using such token definition, and context would capture, we focus on the ``Smart Lab - UNSW'' trace, and we extract a number of protocol fields (e.g., 
ip.src\_host, ip.dst\_host, udp.srcport, udp.dstport, tcp.srcport, tcp.dstport, icmp.type, dns.qry.name, http.server, http.request.method, 
http.request.uri, http.host, ssl.handshake.ciphersuite, http.user\_agent, etc.) using tshark. We create sequences of $N$ tokens, and learn contextual embeddings through the procedure described in Section~\ref{sec:norbert}.

\textbf{Preliminary results:} Similarly to the analysis in Section~\ref{sec:nn}, in order to verify the quality of the learned embeddings, given a token, we retrieve its nearest neighbor. For example, considering the token ``80'', the closest neighbor is ``443''. Destination port 80 represents HTTP, while destination port 443 represents HTTPS. Those results are therefore according to intuition. Next, considering the token ``49199'' which is the code for the ciphersuite ``ECDHE + RSA authentication	AES-128 GCM	SHA-256'', its closest neighbor is token ``49200'' which is the code for ciphersuite ``ECDHE + RSA authentication	AES-256 GCM	SHA-384''. Those two codes represent the same algorithms, and differ only in the keys' lengths. As such, the proximity of these codes is again according to intuition. In summary, the preliminary results are  promising, indicating that the embeddings can learn meaningful relationships for different network categorical variables, though downstream evaluation tasks are left to future work.

\section{Conclusion}
We have adapted the NLP pre-training technique and associated deep model BERT to learn semantically meaningful numerical representations (embeddings) for Fully Qualified Domain Names (FQDNs) used in communication networks. We demonstrate that applying the above approach allows embeddings to capture the semantic information behind FQDNs, and significantly improves deep models’ generalization performance. A series of experiments were run to show the benefits of using BERT and deriving contextual embeddings for FQDNs.
We further analyze the learned embeddings, and present (unsupervised) tasks network administrators could perform thanks to these novel embeddings (e.g., analysis of user behaviors). Finally, we discuss approaches for learning embeddings that capture semantic relationships for network categorical values, other than FQDNs, and present preliminary results.


\bibliographystyle{ieeetr}
\bibliography{reference}

\begin{thebibliography}{10}

\bibitem{8713803}
S.~Rezaei and X.~Liu, ``Deep learning for encrypted traffic classification: An
  overview,'' {\em IEEE Communications Magazine}, vol.~57, no.~5, pp.~76--81,
  2019.

\bibitem{lotfollahi2020deep}
M.~Lotfollahi, M.~Jafari~Siavoshani, R.~Shirali Hossein~Zade, and M.~Saberian,
  ``Deep packet: A novel approach for encrypted traffic classification using
  deep learning,'' {\em Soft Computing}, vol.~24, no.~3, pp.~1999--2012, 2020.

\bibitem{10.1145/3341302.3342221}
E.~Liang, H.~Zhu, X.~Jin, and I.~Stoica, ``Neural packet classification,'' in
  {\em Proceedings of the ACM Special Interest Group on Data Communication},
  SIGCOMM '19, (New York, NY, USA), p.~256–269, Association for Computing
  Machinery, 2019.

\bibitem{unsw}
A.~Sivanathan, D.~Sherratt, H.~H. Gharakheili, A.~Radford, C.~Wijenayake,
  A.~Vishwanath, and V.~Sivaraman, ``{Characterizing and Classifying IoT
  Traffic in Smart Cities and Campuses},'' in {\em {IEEE Infocom Workshop Smart
  Cities and Urban Computing}}, 2017.

\bibitem{9151339}
H.~Guo and J.~Heidemann, ``Detecting iot devices in the internet,'' {\em
  IEEE/ACM Transactions on Networking}, vol.~28, no.~5, pp.~2323--2336, 2020.

\bibitem{8440758}
A.~Sivanathan, H.~H. Gharakheili, F.~Loi, A.~Radford, C.~Wijenayake,
  A.~Vishwanath, and V.~Sivaraman, ``Classifying iot devices in smart
  environments using network traffic characteristics,'' {\em IEEE Transactions
  on Mobile Computing}, vol.~18, no.~8, pp.~1745--1759, 2019.

\bibitem{10.1145/3005745.3005750}
H.~Mao, M.~Alizadeh, I.~Menache, and S.~Kandula, ``Resource management with
  deep reinforcement learning,'' in {\em Proceedings of the 15th ACM Workshop
  on Hot Topics in Networks}, HotNets '16, (New York, NY, USA), p.~50–56,
  Association for Computing Machinery, 2016.

\bibitem{10.1145/2486001.2486020}
K.~Winstein and H.~Balakrishnan, ``Tcp ex machina: Computer-generated
  congestion control,'' in {\em Proceedings of the ACM SIGCOMM 2013 Conference
  on SIGCOMM}, SIGCOMM '13, (New York, NY, USA), p.~123–134, Association for
  Computing Machinery, 2013.

\bibitem{8790762}
W.~Li, H.~Zhang, S.~Gao, C.~Xue, X.~Wang, and S.~Lu, ``Smartcc: A reinforcement
  learning approach for multipath tcp congestion control in heterogeneous
  networks,'' {\em IEEE Journal on Selected Areas in Communications}, vol.~37,
  no.~11, pp.~2621--2633, 2019.

\bibitem{8668690}
X.~Nie, Y.~Zhao, Z.~Li, G.~Chen, K.~Sui, J.~Zhang, Z.~Ye, and D.~Pei, ``Dynamic
  tcp initial windows and congestion control schemes through reinforcement
  learning,'' {\em IEEE Journal on Selected Areas in Communications}, vol.~37,
  no.~6, pp.~1231--1247, 2019.

\bibitem{10.1145/3512798.3512815}
C.~Tessler, Y.~Shpigelman, G.~Dalal, A.~Mandelbaum, D.~Haritan~Kazakov,
  B.~Fuhrer, G.~Chechik, and S.~Mannor, ``Reinforcement learning for datacenter
  congestion control,'' {\em SIGMETRICS Perform. Eval. Rev.}, vol.~49,
  p.~43–46, jan 2022.

\bibitem{NIPS1993_4ea06fbc}
J.~Boyan and M.~Littman, ``Packet routing in dynamically changing networks: A
  reinforcement learning approach,'' in {\em Advances in Neural Information
  Processing Systems} (J.~Cowan, G.~Tesauro, and J.~Alspector, eds.), vol.~6,
  Morgan-Kaufmann, 1993.

\bibitem{8851733}
J.~Reis, M.~Rocha, T.~K. Phan, D.~Griffin, F.~Le, and M.~Rio, ``Deep neural
  networks for network routing,'' in {\em 2019 International Joint Conference
  on Neural Networks (IJCNN)}, pp.~1--8, 2019.

\bibitem{8701570}
Z.~Mammeri, ``Reinforcement learning based routing in networks: Review and
  classification of approaches,'' {\em IEEE Access}, vol.~7, pp.~55916--55950,
  2019.

\bibitem{9614298}
Y.~Chen and G.~Casale, ``Deep learning models for automated identification of
  scheduling policies,'' in {\em 2021 29th International Symposium on Modeling,
  Analysis, and Simulation of Computer and Telecommunication Systems
  (MASCOTS)}, pp.~1--8, 2021.

\bibitem{10.1145/2834050.2834095}
T.~Yu, V.~Sekar, S.~Seshan, Y.~Agarwal, and C.~Xu, ``Handling a trillion
  (unfixable) flaws on a billion devices: Rethinking network security for the
  internet-of-things,'' in {\em Proceedings of the 14th ACM Workshop on Hot
  Topics in Networks}, HotNets-XIV, (New York, NY, USA), Association for
  Computing Machinery, 2015.

\bibitem{wagner1996analysis}
D.~Wagner, B.~Schneier, {\em et~al.}, ``Analysis of the ssl 3.0 protocol,'' in
  {\em The Second USENIX Workshop on Electronic Commerce Proceedings}, vol.~1,
  pp.~29--40, 1996.

\bibitem{ou2012exploiting}
Z.~Ou, H.~Zhuang, J.~K. Nurminen, A.~Yl{\"a}-J{\"a}{\"a}ski, and P.~Hui,
  ``Exploiting hardware heterogeneity within the same instance type of amazon
  $\{$EC2$\}$,'' in {\em 4th USENIX Workshop on Hot Topics in Cloud Computing
  (HotCloud 12)}, 2012.

\bibitem{10.1145/78952.78955}
K.~K. Ramakrishnan and R.~Jain, ``A binary feedback scheme for congestion
  avoidance in computer networks,'' {\em ACM Trans. Comput. Syst.}, vol.~8,
  p.~158–181, may 1990.

\bibitem{devlin2018bert}
J.~Devlin, M.-W. Chang, K.~Lee, and K.~Toutanova, ``Bert: Pre-training of deep
  bidirectional transformers for language understanding,'' {\em arXiv preprint
  arXiv:1810.04805}, 2018.

\bibitem{pennington-etal-2014-glove}
J.~Pennington, R.~Socher, and C.~Manning, ``{G}lo{V}e: Global vectors for word
  representation,'' in {\em Proceedings of the 2014 Conference on Empirical
  Methods in Natural Language Processing ({EMNLP})}, (Doha, Qatar),
  pp.~1532--1543, Association for Computational Linguistics, Oct. 2014.

\bibitem{mikolov2013efficient}
T.~Mikolov, K.~Chen, G.~Corrado, and J.~Dean, ``Efficient estimation of word
  representations in vector space,'' {\em arXiv preprint arXiv:1301.3781},
  2013.

\bibitem{6289079}
M.~Schuster and K.~Nakajima, ``Japanese and korean voice search,'' in {\em 2012
  IEEE International Conference on Acoustics, Speech and Signal Processing
  (ICASSP)}, pp.~5149--5152, 2012.

\bibitem{sennrich-etal-2016-neural}
R.~Sennrich, B.~Haddow, and A.~Birch, ``Neural machine translation of rare
  words with subword units,'' in {\em Proceedings of the 54th Annual Meeting of
  the Association for Computational Linguistics (Volume 1: Long Papers)},
  (Berlin, Germany), pp.~1715--1725, Association for Computational Linguistics,
  Aug. 2016.

\bibitem{NIPS2017_3f5ee243}
A.~Vaswani, N.~Shazeer, N.~Parmar, J.~Uszkoreit, L.~Jones, A.~N. Gomez, L.~u.
  Kaiser, and I.~Polosukhin, ``Attention is all you need,'' in {\em Advances in
  Neural Information Processing Systems} (I.~Guyon, U.~V. Luxburg, S.~Bengio,
  H.~Wallach, R.~Fergus, S.~Vishwanathan, and R.~Garnett, eds.), vol.~30,
  Curran Associates, Inc., 2017.

\bibitem{bahdanau2014neural}
D.~Bahdanau, K.~Cho, and Y.~Bengio, ``Neural machine translation by jointly
  learning to align and translate,'' {\em arXiv preprint arXiv:1409.0473},
  2014.

\bibitem{lan2019albert}
Z.~Lan, M.~Chen, S.~Goodman, K.~Gimpel, P.~Sharma, and R.~Soricut, ``Albert: A
  lite bert for self-supervised learning of language representations,'' {\em
  arXiv preprint arXiv:1909.11942}, 2019.

\bibitem{liu2019roberta}
Y.~Liu, M.~Ott, N.~Goyal, J.~Du, M.~Joshi, D.~Chen, O.~Levy, M.~Lewis,
  L.~Zettlemoyer, and V.~Stoyanov, ``Roberta: A robustly optimized bert
  pretraining approach,'' {\em arXiv preprint arXiv:1907.11692}, 2019.

\bibitem{10.5555/2028067.2028094}
M.~Antonakakis, R.~Perdisci, W.~Lee, N.~Vasiloglou, and D.~Dagon, ``Detecting
  malware domains at the upper dns hierarchy,'' in {\em Proceedings of the 20th
  USENIX Conference on Security}, SEC'11, (USA), p.~27, USENIX Association,
  2011.

\bibitem{10.1145/2584679}
L.~Bilge, S.~Sen, D.~Balzarotti, E.~Kirda, and C.~Kruegel, ``Exposure: A
  passive dns analysis service to detect and report malicious domains,'' {\em
  ACM Trans. Inf. Syst. Secur.}, vol.~16, apr 2014.

\bibitem{10.5555/2460416.2460425}
S.~Campbell, S.~Chan, and J.~R. Lee, ``Detection of fast flux service
  networks,'' in {\em Proceedings of the Ninth Australasian Information
  Security Conference - Volume 116}, AISC '11, (AUS), p.~57–66, Australian
  Computer Society, Inc., 2011.

\bibitem{Holz:2008}
T.~Holz, C.~Gorecki, K.~Rieck, and F.~C. Freiling, ``Measuring and {D}etecting
  {F}ast-{F}lux {S}ervice {N}etworks,'' in {\em Symp NDSS}, 2008.

\bibitem{Perdisci:2009}
R.~Perdisci, I.~Corona, D.~Dagon, and W.~Lee, ``Detecting {M}alicious {F}lux
  {S}ervice {N}etworks {T}hrough {P}assive {A}nalysis of {R}ecursive {DNS}
  {T}races,'' in {\em Annual Computer Security Applications Conference}, 2009.

\bibitem{10.1145/1879141.1879148}
S.~Yadav, A.~K.~K. Reddy, A.~N. Reddy, and S.~Ranjan, ``Detecting
  algorithmically generated malicious domain names,'' in {\em Proceedings of
  the 10th ACM SIGCOMM Conference on Internet Measurement}, IMC '10, (New York,
  NY, USA), p.~48–61, Association for Computing Machinery, 2010.

\bibitem{singh-etal-2019-bert}
J.~Singh, B.~McCann, R.~Socher, and C.~Xiong, ``{BERT} is not an interlingua
  and the bias of tokenization,'' in {\em Proceedings of the 2nd Workshop on
  Deep Learning Approaches for Low-Resource NLP (DeepLo 2019)}, (Hong Kong,
  China), pp.~47--55, Association for Computational Linguistics, Nov. 2019.

\bibitem{kingma2014adam}
D.~P. Kingma and J.~Ba, ``Adam: A method for stochastic optimization,'' {\em
  arXiv preprint arXiv:1412.6980}, 2014.

\bibitem{Bro}
V.~Paxson, ``{Bro: a System for Detecting Network Intruders in Real-Time},''
  {\em Computer Networks}, vol.~31, no.~23-24, pp.~2435--2463, 1999.

\bibitem{cho-etal-2014-learning}
K.~Cho, B.~van Merri{\"e}nboer, C.~Gulcehre, D.~Bahdanau, F.~Bougares,
  H.~Schwenk, and Y.~Bengio, ``Learning phrase representations using {RNN}
  encoder{--}decoder for statistical machine translation,'' in {\em Proceedings
  of the 2014 Conference on Empirical Methods in Natural Language Processing
  ({EMNLP})}, (Doha, Qatar), pp.~1724--1734, Association for Computational
  Linguistics, Oct. 2014.

\bibitem{10.5555/3045390.3045502}
Y.~Gal and Z.~Ghahramani, ``Dropout as a bayesian approximation: Representing
  model uncertainty in deep learning,'' in {\em Proceedings of the 33rd
  International Conference on International Conference on Machine Learning -
  Volume 48}, ICML'16, p.~1050–1059, JMLR.org, 2016.

\bibitem{10.5555/3295222.3295387}
B.~Lakshminarayanan, A.~Pritzel, and C.~Blundell, ``Simple and scalable
  predictive uncertainty estimation using deep ensembles,'' in {\em Proceedings
  of the 31st International Conference on Neural Information Processing
  Systems}, NIPS'17, (Red Hook, NY, USA), p.~6405–6416, Curran Associates
  Inc., 2017.

\bibitem{liang2017enhancing}
S.~Liang, Y.~Li, and R.~Srikant, ``Enhancing the reliability of
  out-of-distribution image detection in neural networks,'' {\em arXiv preprint
  arXiv:1706.02690}, 2017.

\bibitem{10.5555/3327757.3327819}
K.~Lee, K.~Lee, H.~Lee, and J.~Shin, ``A simple unified framework for detecting
  out-of-distribution samples and adversarial attacks,'' in {\em Proceedings of
  the 32nd International Conference on Neural Information Processing Systems},
  NIPS'18, (Red Hook, NY, USA), p.~7167–7177, Curran Associates Inc., 2018.

\bibitem{NIPS2002_6150ccc6}
G.~E. Hinton and S.~Roweis, ``Stochastic neighbor embedding,'' in {\em Advances
  in Neural Information Processing Systems} (S.~Becker, S.~Thrun, and
  K.~Obermayer, eds.), vol.~15, MIT Press, 2002.

\bibitem{rfc2616}
H.~Nielsen, J.~Mogul, L.~M. Masinter, R.~T. Fielding, J.~Gettys, P.~J. Leach,
  and T.~Berners-Lee, ``{Hypertext Transfer Protocol -- HTTP/1.1}.'' RFC 2616,
  June 1999.

\bibitem{rfc7252}
Z.~Shelby, K.~Hartke, and C.~Bormann, ``{The Constrained Application Protocol
  (CoAP)}.'' RFC 7252, June 2014.

\bibitem{rfc4254}
T.~Ylonen and C.~Lonvick, ``{The Secure Shell (SSH) Connection Protocol}.'' RFC
  4254 (Proposed Standard), January 2006.

\bibitem{rfc5246}
E.~Rescorla and T.~Dierks, ``{The Transport Layer Security (TLS) Protocol
  Version 1.2}.'' RFC 5246, Aug. 2008.

\bibitem{10.5555/1202316}
A.~Orebaugh, G.~Ramirez, J.~Beale, and J.~Wright, {\em Wireshark \& Ethereal
  Network Protocol Analyzer Toolkit}.
\newblock Syngress Publishing, 2007.

\end{thebibliography}

\end{document}